\documentclass[9pt,twocolumn,twoside]{osajnl}
\journal{ol} 

\usepackage{bm}
\usepackage{amsmath}
\usepackage{amssymb}
\usepackage{graphicx}
 \usepackage{esint}

\setboolean{shortarticle}{true}

\title{Robustness of quantum Fourier transform interferometry}

\author[1]{Bogdan~Opanchuk}
\author[1,2]{Laura~Rosales-Z\'arate}
\author[1,3]{Margaret~D.~Reid}
\author[1,3,*]{Peter~D.~Drummond}

\affil[1]{Centre for Quantum and Optical Science, Swinburne University of Technology,
Melbourne 3122, Australia}
\affil[2]{Centro de Investigaciones en Optica A.C., Leon, Guanajuato 37150, Mexico}

\affil[3]{Institute of Theoretical Atomic, Molecular and Optical Physics (ITAMP),Harvard
University, Cambridge, Massachusetts, USA.}

\affil[*]{Corresponding author: peterddrummond@gmail.com}

\begin{abstract}
We analyse the effect of decoherence
and noise on quantum Fourier transform interferometry, in which a boson sampling photonic network
is used to measure optical phase gradients. 
This novel type of metrology is shown to be robust against phase decoherence.
One can also measure gradients using lower order correlations without
substantial degradation. Our results involve estimation
of up to a $100\times100$ matrix permanent.
\end{abstract}

\setboolean{displaycopyright}{true}

\begin{document}

\maketitle

Linear photonic networks employing a quantum computing technique called
Boson Sampling~\cite{AaronsonArkhipov2013LV, wang2017high} can be used to implement
quantum Fourier transform (QFT) interferometers~\cite{Motes2015_PRL114,Su2017Multiphoton}.
These use high order quantum correlations to measure phase gradients,
 with potential applications to high precision metrology. Here
we focus on the robustness of the technique against phase-noise and imperfect correlations.

To understand the performance of a real gradient measurement interferometer
one needs to include decoherence. The object being measured will not
have a completely uniform phase gradient. This is a potential advantage
for the QFT interferometer, as it intrinsically uses multiple channels
to analyse the gradient. A multiple channel interferometer is able 
to average over any introduced phase noise, thus producing a robust
measurement even in the presence of external noise.

An important question is
whether the scaling of an ideal interferometer can be translated to
a real world case with phase noise and loss, leading to imperfect correlations. 
However, in the case of a large QFT network,
the calculation involves evaluating a large matrix permanent. Exact
methods are generally not feasible for $M\times M$ QFT interferometers
above $M=50$~\cite{Wu2016arXiv160605836}, similar to limits for
computationally generating just one random number state sample \cite{neville2017classical}
with a permanental distribution. 

We analyze much larger QFT interferometers than this, with up to $M=100$
channels. We calculate the effects of noise using  
coherent state expansions. These results demonstrate that the QFT interferometer
is remarkably robust against phase noise, with only a very gradual degradation
of peak visibility with phase-noise, not a catastrophic failure when
imperfections occur. 

Calculating this exactly for a general unitary matrix is exponentially
hard for large matrices, as it involves the square of a matrix permanent
which is in the complexity class $\#P$~\cite{Valiant1979}. Our methods solve this problem using phase-space representations from quantum optics, and
 far exceed previous size limits on permanent squared calculations.
Our methods can represent any input state or output
measurement. This is essential for understanding how this technology
can be used with imperfect sources~\cite{He2016ScalableBS},
detectors, or for different applications~\cite{OlsonLinearOptQM}.

To illustrate this issue, we also calculate the QFT correlation peak
for correlation functions of much less than optimal order. Such low-order
correlations are another realistic requirement in a practical device, owing to inevitable detector and channel losses.
Here the technique is again surprisingly robust. Lowering the correlation
order has relatively little effect on the peak width. Generating pure single-photon
states in large numbers of channels simultaneously is difficult. However, we conjecture
that the robustness of the method against noise and correlation order
may allow imperfect input states as well. 

\paragraph*{Complex P-representations.}

A boson sampling interferometer uses a multichannel photonic network to obtain interference fringes with
nonclassical input states, typically number states. The output is  nontraditional, as the interferometer measures coincidence counts between single-particle counts in all of the many output channels. These are
required to be counted simultaneously for a count contributing to a fringe measurement. As a result,
one cannot use classical optical theoretical methods. 

We use the complex $P$-representation~\cite{Drummond_Gardiner_PositivePRep, ScalingPerm, opanchuk2018simulating}
for the input and the output states of the boson sampling interferometer. This is a more powerful phase-space method than the
 classical phase-space methods considered previously for this problem~\cite{RahimiKeshari:2016}, because it
expands the quantum density matrix $\hat{\rho}$ using off-diagonal coherent
state projectors. This extends methods developed by Glauber~\cite{Glauber_1963_P-Rep}, resulting in a compact and efficient representation.  While fully quantum-mechanical,
the technique employs a coherent state basis set of amplitudes with a rather classical behaviour. 
It uses contours $C$ enclosing the origins of the complex planes
corresponding to the variables $\alpha_{k}$, $\beta_{k}$:
\begin{equation}
\hat{\rho}^{(\mathrm{in})}=\oiint_{C}P(\bm{\alpha},\bm{\beta})\hat{\Lambda}\left(\bm{\alpha},\bm{\beta}\right)\mathrm{d}\bm{\alpha}\mathrm{d}\bm{\beta}\,.
\end{equation}
The projection operator $\hat{\Lambda}\left(\bm{\alpha},\bm{\beta}\right)=\left\Vert \bm{\alpha}\right\rangle \left\langle \bm{\beta}^{*}\right\Vert /\left\langle \bm{\beta}^{*}\right\Vert \left.\bm{\alpha}\right\rangle $
is defined in terms of $M$-mode un-normalized Bargmann-Glauber coherent
states $\left\Vert \bm{\alpha}\right\rangle $, where
\begin{equation}
\left\Vert \bm{\alpha}\right\rangle \equiv\prod_{k=1}^{M} \sum_{n_{k}}  \frac{\alpha_{k}^{n_{k}}}{\sqrt{n_{k}!}} \left|n_{k}\right\rangle \,.
\end{equation}
In
this work we use circular contours of radius $r$, where $r$ can
be varied, and this changes the efficiency of the representation. 

A complex $P$ function exists for any density matrix with bounded
support~\cite{Drummond_Gardiner_PositivePRep}. Here we consider
the initial state to be a pure multimode state where each mode $k$
is a number state with $n_{k}$ bosons. In the experiments at most
one photon is input per mode, so $n_{k}=0,1$.

An application of this method to boson sampling \cite{opanchuk2018simulating}
is summarized briefly here. The input density matrix $\hat{\rho}^{\mathrm{(in)}}$
is transformed to the output density matrix $\hat{\rho}^{\mathrm{(out)}}$
by a linear network, described by a unitary matrix $\bm{U}$. Losses
can be treated by adding modes to serve as noise channels, but here
we consider the unitary case. The effect of the network matrix on
the phase-space variables is a simple matrix product: $\bm{\alpha}^{(\mathrm{out})},\,\bm{\beta}^{(\mathrm{out})}=\bm{U}\bm{\alpha},\,\bm{U}^{*}\bm{\beta}$~\cite{drummond2014quantum}.
The output density matrix is
\begin{equation}
\hat{\rho}^{(\mathrm{out})}=\mathfrak{R}\oiint_{C}P(\bm{\alpha},\bm{\beta})\hat{\Lambda}\left(\bm{U}\bm{\alpha},\,\bm{U}^{*}\bm{\beta}\right)\mathrm{d}\bm{\alpha}\mathrm{d}\bm{\beta}\,.
\end{equation}

The observable in photonic experiments is a normally ordered correlation
of output modes in a set $\sigma^{\prime}$;
\begin{eqnarray}
\left\langle \prod_{k\in\sigma^{\prime}}\hat{n}_{k}\right\rangle _{\mathrm{Q}} & = & \oiint_{C}P(\bm{\alpha},\bm{\beta})\prod_{k\in\sigma^{\prime}}n_{k}^{(\mathrm{out})}(\bm{\alpha},\bm{\beta})\mathrm{d}\bm{\alpha}\mathrm{d}\bm{\beta}\nonumber \\
 & \equiv & \left\langle \prod_{k\in\sigma^{\prime}}n_{k}^{(\mathrm{out})}(\bm{\alpha},\bm{\beta})\right\rangle _{\mathrm{P}}.
\end{eqnarray}
Here the classical output number variable is defined as $n_{k}^{(\mathrm{out})}(\bm{\alpha},\bm{\beta})=\alpha_{k}^{(\mathrm{out})}\beta_{k}^{(\mathrm{out})}$.

As the distribution is not unique, one can choose different representations
of the input state. In this work we use two kinds of representations
tailored for different scenarios, described below: a continuous sampling
method (VCP) and a discrete method (QCP). Which one is preferred depends
on the correlation being calculated, as we explain in detail next.

\paragraph*{Continuous sampling method (VCP).}

We  first consider a continuous analytic complex $P$-distribution for the
initial state that is input to an arbitrary photonic network, 
\begin{equation}
P\left(\bm{\alpha},\bm{\beta}\right)=\prod_{k=1}^{M}P_{k}\left(\alpha_{k},\beta_{k}\right)\,.
\end{equation}
$P_{k}$ is the single-mode distribution for the $k$-th input channel, with $P_k = \delta\left(\alpha\right)\delta\left(\beta\right) $ for $\ n_{k}=0$, and otherwise:
\begin{equation}
P_{k}\left(\alpha,\beta\right)=
\left(\frac{n_{k}!}{2\pi\mathrm{i}}\right)^{2}\frac{e^{\alpha\beta}}{(\alpha\beta)^{n_{k}+1}} .
\end{equation}

Here $n_{k}$ is the number of photons in the input mode $k$, and
for modes where $n_{k}=0$ we shrink the contours to the origin of
the complex plane. For nonzero boson number inputs we use a circular
contour of radius $r$ with polar coordinates, so that $\alpha_{k}=rz_{k}=r\exp\left(\mathrm{i}\phi_{k}^{(\alpha)}\right)$
and $\beta_{k}=r\tilde{z}_{k}^{*}=r\exp\left(-\mathrm{i}\phi_{k}^{(\beta)}\right)$.
The radius $r$ is chosen to minimize the sampling error. 

The phase variables are simplified on defining $\phi_{k}=\left(\phi_{k}^{(\alpha)}+\phi_{k}^{(\beta)}\right)/2$
and $\theta_{k}=\phi_{k}^{(\alpha)}-\phi_{k}^{(\beta)}$, where $\phi_{k}$
is the classical phase, and $\theta_{k}$ is a nonclassical phase
which only exists when the quantum state is nonclassical,
so $\hat{\rho}=\prod_{k=1}^{M}J_{k}\left(\phi_{k},\theta_{k}\right)\hat{\Lambda}\left(\alpha_{k},\beta_{k}\right),$
where $J_{k}=1$ for $n_{k}=0$, and otherwise: 
\begin{equation}
J_{k}=\frac{(n_{k}!)^{2}}{4\pi^{2}r^{2n_{k}}}\int_{-\pi}^{\pi}\mathrm{d}\theta_{k}\int_{-\pi}^{\pi}\mathrm{d}\phi_{k}e^{\left(r^{2}\cos\theta_{k}+\mathrm{i}\left(r^{2}\sin\theta_{k}-n_k\theta_{k}\right)\right)} .
\end{equation}
We use random probabilistic sampling for the real part, and call the
distribution a circular von Mises complex-P distribution (VCP)~\cite{BookvonMisesCh3}.
Once the phase angle is chosen,  the imaginary part becomes a complex
weight. The sampled probability distribution for modes with nonzero
inputs is
\begin{equation}
P_{k}\left(\phi_{k},\theta_{k}\right)=\frac{1}{2\pi I_{0}\left(r^{2}\right)}\exp\left(r^{2}\cos\theta_{k}\right),\quad n_{k}>0,
\end{equation}
where $I_{0}\left(x\right)$ is the modified Bessel function of the
first kind of order $0$, $\theta_{k}\in[-\pi,\pi)$, and $\phi_{k}\in[-\pi,\pi)$.
This corresponds to sampling the variables separately as $\phi_{k}=\mathcal{U}\left(-\pi,\pi\right)$,
and $\theta_{k}=\mathcal{VM}\left(0,r^{2}\right)$, where $\mathcal{U}$
is the uniform distribution, and $\mathcal{VM}$ is the circular von
Mises distribution~\cite{BookvonMisesCh3}. For each sample we calculate
the phase-space variables as $\alpha_{k}=r\exp\left(\mathrm{i}\left(\phi_{k}+\theta_{k}/2\right)\right)$,
$\beta_{k}=r\exp\left(\mathrm{i}\left(\phi_{k}-\theta_{k}/2\right)\right)$.
For factors where $n_{k}=0$ we use the optimal distribution $P_{k}\left(\alpha_{k},\beta_{k}\right)=\delta\left(\alpha_{k}\right)\delta\left(\beta_{k}\right)$
instead, that is, take $\alpha_{k}=\beta_{k}=0$.

The corresponding complex weights are $\Omega=\prod_{k=1}^{M}\Omega_{k}$
where $\Omega_{k}=1$ for $n_{k}=0$, and otherwise: 
\begin{equation}
\Omega_{k}=\frac{\left(n_{k}!\right)^{2}I_{0}\left(r^{2}\right)}{r^{2n_{k}}}\exp\left(\mathrm{i}\left(r^{2}\sin\theta_{k}-n_{k}\theta_{k}\right)\right),
\end{equation}

These are used to calculate any moment $f(\bm{\alpha},\bm{\beta})$
in conjunction with the drawn samples of $\bm{\alpha}$ and $\bm{\beta}$
as
\begin{equation}
\langle f(\bm{\alpha},\bm{\beta})\rangle=\frac{1}{L}\sum_{j=1}^{L}\Omega\left(\bm{\alpha}^{(j)},\bm{\beta}^{(j)}\right)f\left(\bm{\alpha}^{(j)},\bm{\beta}^{(j)}\right),
\end{equation}
where $L$ denotes the total number of samples, and $\bm{\alpha}^{(j)}$
and $\bm{\beta}^{(j)}$ is the $j$-th sample of a pair of random
phase-space coordinates.

\paragraph*{Discrete sampling method (QCP).}

Suppose that the initial photon number is bounded, for example with
a fixed input boson number. We now introduce the discrete or qudit
complex P-representations (QCP)~\cite{Drummond2016-coherent}. This
is useful in the limit of $r\rightarrow0$, which is expanded using
$d$ coherent phases distributed on a circle. It is equivalent to
a $d$-dimensional qudit, with initial occupation numbers in each
mode of $n=0,\ldots d-1$. This unifies quantum optics~\cite{Drummond_Gardiner_PositivePRep}
with discrete sampling permanent approximation methods~\cite{gurvits2002deterministic}. 

Here we limit the possible values of the phase-space variables $\alpha_{j}$
and $\beta_{j}$ to a finite number of values defined by two vectors
$\bm{q}$ and $\tilde{\bm{q}}$ of discrete coherent amplitudes so
that $\alpha_{j}\left(q_{j}\right)=rz^{q_{j}},\quad q{}_{j}=0,\ldots d-1,$
and $\beta_{j}\left(\tilde{q}_{j}\right)=rz^{-\tilde{q}{}_{j}},\quad\tilde{q}{}_{j}=0,\ldots d-1,$
where $z=\exp\left(2\pi\mathrm{i}/d\right)$. The density matrix is
expanded in coherent states with a discrete summation, $\hat{\rho}=\sum_{\boldsymbol{q},\tilde{\boldsymbol{q}}}P(\boldsymbol{q},\tilde{\boldsymbol{q}})\hat{\Lambda}(\boldsymbol{q},\tilde{\boldsymbol{q}})/d^{2M}\,,$where
$\hat{\Lambda}(\boldsymbol{q},\tilde{\boldsymbol{q}})=\left\Vert \vphantom{\bm{\beta}^{*}(\tilde{\mathbf{q}})}\bm{\alpha}(\boldsymbol{q})\right\rangle \left\langle \bm{\beta}^{*}(\tilde{\boldsymbol{q}})\right\Vert $.
The coherent states have unit norm in the limit of $r\rightarrow0$.
Using this expansion, a complex qudit P-function $P_{Q}$ always exists
for a given input density matrix $\hat{\rho}$, where:
\begin{equation}
P_{Q}(\boldsymbol{q},\tilde{\boldsymbol{q}})=\sum_{\bm{n},\bm{m}}\left\langle \bm{m}\right|\hat{\rho}\left|\bm{n}\right\rangle \prod_{j=1}^{M}\frac{\sqrt{n_{j}!m_{j}!}}{r^{n_{j}+m_{j}}}\mathrm{e}^{\frac{2\pi\mathrm{i}}{d}(n_{j}\tilde{q}_{j}-m_{j}q_{j})}\,.
\end{equation}
The result is like a path integral, except that the distribution is
a discrete sum. This can be verified as a solution, by noting that
$\sum_{\boldsymbol{q}}\mathrm{e}^{\mathrm{i}\bm{q}\cdot\left(\bm{n}-\bm{m}\right)\phi}=d^{M}\delta_{\bm{n}-\bm{m}}.$
In the one mode, one boson case, the P-function is the product of
two complex variables with a radial factor, so that $P_{\mathrm{Q}}(q,\tilde{q})=\tilde{z}^{\tilde{q}}\left(z^{q}\right)^{*}/r^{2}\,.$

In the qudit case, for a single-mode Hilbert space dimension $d$,
we consider $d$ random discrete phases. Here $\phi_{k}$ can have
values of $(0,\ldots d-1)\times\phi$, with $\phi=2\pi/d$, allowing
occupation numbers for mode $k$ up to $n_{k}=0,\ldots d-1$. We note
that such discrete sampling reduces to continuous sampling in the
limit of $d\rightarrow\infty$. In the simplest binary, or qubit,
case where $n_{k}=0,1$ we take the minimum value of $d=2$, so that
$\alpha_{k}=\pm r$.

The photon number phase-space variable, for an input of single bosons
into a subset $\sigma$ of size $N$ of input modes, is 
\begin{equation}
n_{k}^{\mathrm{(out)}}(q,\tilde{q})=r^{2}\left(\sum_{j\in\sigma}U_{kj}z^{q_{j}}\right)\left(\sum_{j^{\prime}\in\sigma}U_{kj^{\prime}}z^{\tilde{q}_{j^{\prime}}}\right)^{*}.
\end{equation}
As a result, after including the complex $P$-function weights, an
$N$-th order output correlation is
\begin{eqnarray}
\left\langle \prod_{k\in\sigma^{\prime}}\hat{n}_{k}\right\rangle _{\mathrm{Q}} & = & \left|\frac{1}{d^{M}}\sum_{\boldsymbol{q}}\left\{ \prod_{i\in\sigma}z^{-q_{i}}\prod_{k\in\sigma^{\prime}}\left(\sum_{j\in\sigma}U_{kj}z^{q_{j}}\right)\right\} \right|^{2}.\nonumber \\
 & = & \left|\mathrm{perm}\left[U\left(\sigma^{\prime},\sigma\right)\right]\right|^{2}
\end{eqnarray}
Here $\mathrm{perm}$ is the matrix permanent. If the order
of the correlation is equal to the number of input bosons,
the $r$ factors cancel.

As expected~\cite{Aaronson2011}, this is the square of the permanent
of the sub-matrix of $U$ with rows in $\sigma^{\prime}$ and columns
in $\sigma$, which we call $U\left(\sigma^{\prime},\sigma\right)$.
After summation on the $\boldsymbol{q}$ indices, the only terms that
survive involve products of distinct permutations of the matrix indices,
which is the permanent. This is evaluated computationally by taking
randomly chosen integers $\left(\boldsymbol{q},\tilde{\boldsymbol{q}}\right)$,
and averaging over many samples of these random phases.

In order to do this, we define a function 
\begin{equation}
p\left(\sigma,\sigma^{\prime},\bm{q}\right)=\prod_{i\in\sigma}z^{-q_{i}}\prod_{k\in\sigma^{\prime}}\left(\sum_{j\in\sigma}U_{kj}z^{q_{j}}\right),
\end{equation}
representing
a single sample of the permanent for the sub-matrix defined by the
set of inputs $\sigma$ and the set of outputs $\sigma^{\prime}$,
with $\bm{q}$ being a vector of random numbers. Then the permanent
of the sub-matrix can be calculated probabilistically as
\begin{eqnarray}
\mathrm{perm}U\left(\sigma,\sigma^{\prime}\right) & \approx & \frac{1}{L}\sum_{j=1}^{L}p\left(\sigma,\sigma^{\prime},\bm{q}^{(j)}\right)\nonumber \\
 & \equiv & \langle p\left(\sigma,\sigma^{\prime},\bm{q}\right)\rangle_{L}.
\end{eqnarray}
To avoid bias in the calculation of the permanent-squared,
we use independent sets of random variables to calculate the expectations
of the permanent and  its conjugate, so that:
\begin{equation}
\left|\mathrm{perm}\left[U\left(\sigma^{\prime},\sigma\right)\right]\right|^{2}\approx\mathfrak{R}\langle p\left(\sigma,\sigma^{\prime},\bm{q}\right)\rangle_{L}\langle p\left(\sigma,\sigma^{\prime},\tilde{\bm{q}}\right)\rangle_{L}.\label{eq:PermS}
\end{equation}

The noise variables factor into two terms which are independent but
conjugate on average. The factored terms give independent estimates
of the permanent and its conjugate, so their product is an unbiased
estimate of the modulus squared. We impose the constraint that the
final result must be real. To estimate sampling errors\cite{opanchuk2018simulating}, we divide the $L$ terms into  $L=L_1 L_2$, with  $L_1 $ subensembles of $L_2$ random terms.

\paragraph*{Quantum Fourier transform interferometry. }

Now we consider the quantum Fourier transform interferometer (QuFTI)
experiment~\cite{Motes2015_PRL114}. This is a novel multi-mode interferometer
that measures gradients of a phase-shift and corresponds to consider
the case where the input state is one photon in each input mode. Here
the permanent of the full matrix is evaluated, so that $M=N$. Motes
\emph{et al}~\cite{Motes2015_PRL114} conjectured that the count rate, $Q^{(\mathrm{conj})}$
is:
\begin{equation}
\begin{split}Q^{(\mathrm{conj})} & =\prod_{j=1}^{M-1}\frac{2j(M-j)\cos(M\phi)+M^{2}-2jM+2j^{2}}{M^{2}}.\end{split}
\label{eq:ExpConjecturePermS}
\end{equation}

\begin{figure}
\begin{centering}
\includegraphics[width=0.9\columnwidth]{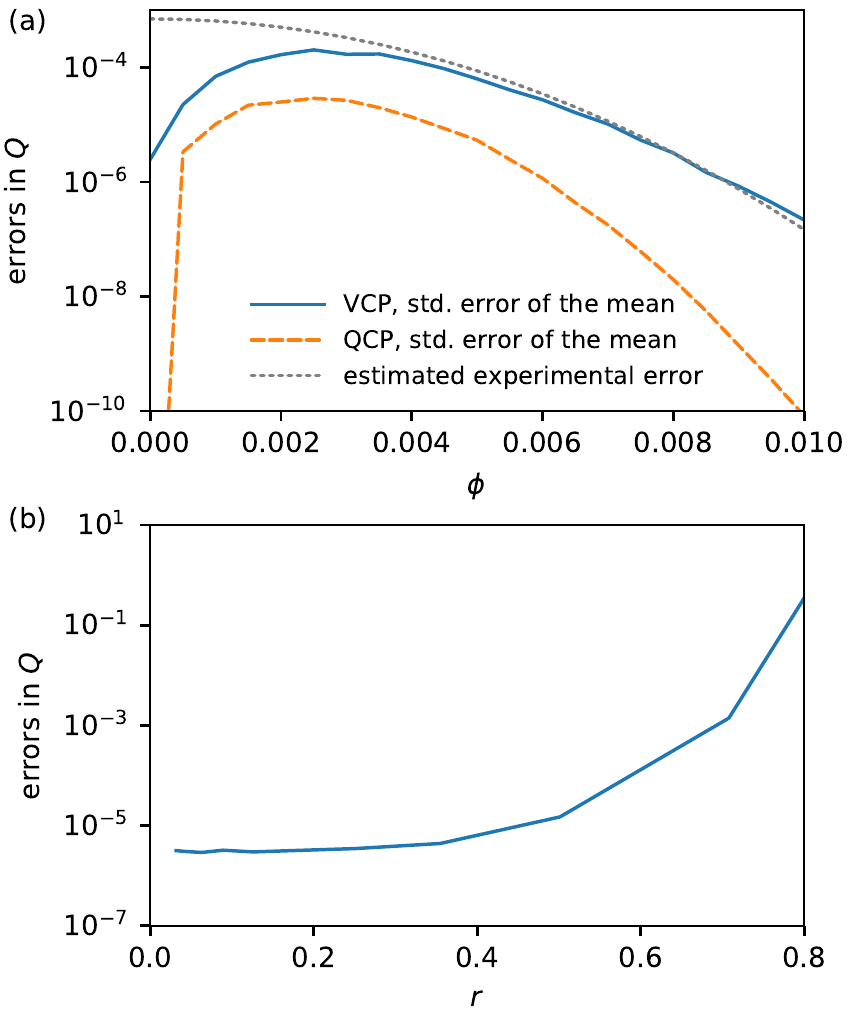}
\par\end{centering}
\caption{\label{fig:max-order}(a) Estimated sampling error in the mean of modulus square
of the permanent $Q$, $\sqrt{\langle Q^{2}\rangle-\langle Q\rangle^{2}}/\sqrt{L_{1}}$,
using two different methods. The solid blue
line corresponds to the VCP with $r=0.1$, while the red dashed line
corresponds to the QCP with $d=2$. The dashed grey line is the estimated
experimental sampling error $\sqrt{Q^{(\mathrm{conj})}/\left(L_{1}L_{2}\right)}$
($L_{1}=200$, $L_{2}=10^{4}$). Here $Q^{(\mathrm{conj})}$ is given
in equation~(\ref{eq:ExpConjecturePermS}). (b) Dependence of the
estimated error on $r$ for the VCP method. Here we have used $M=100$
and $\phi=0.007$, $L_{1}=200$, $L_{2}=10^{5}$.}
\end{figure}

\begin{figure}
\begin{centering}
\includegraphics[width=0.9\columnwidth]{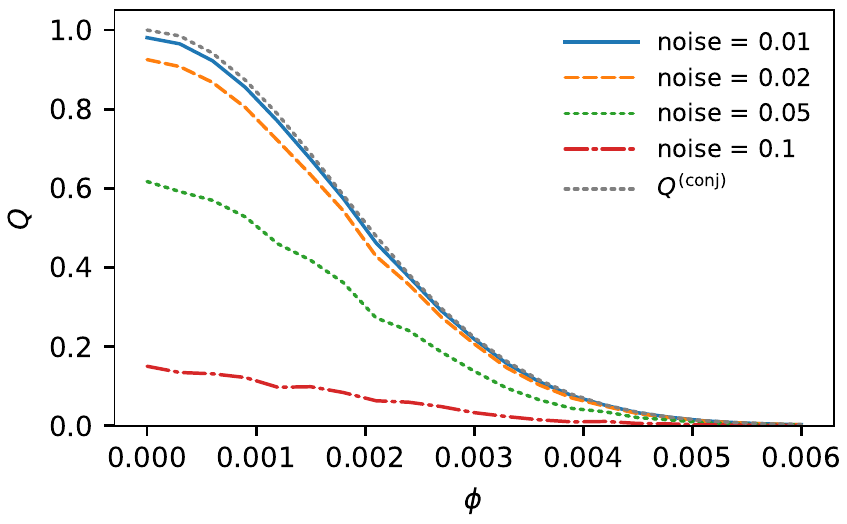}
\par\end{centering}
\caption{\label{fig:noisy}Value of the interferometry correlation $Q$, for
different amounts of noise included in the measured angle. The values
were obtained using QCP with $d=2$. Here we have considered $M=100$,
$L_{1}=200$ ensembles, $L_{2}=10^{4}$, the results are averaged
over $20$ matrices. }
\end{figure}

Fig.~\ref{fig:max-order} shows results obtained for the QuFTI\@.
We confirm the analytic result numerically, with a permanent size
of $100\times100$. Exact numerical methods for permanents are strongly
limited to about $50\times50$ matrices~\cite{Wu2016arXiv160605836},
even with large supercomputers. In Fig.~\ref{fig:max-order}(a) we
evaluate the sampling error in the modulus square of the permanent using both
the VCP and QCP representation with $d=2$. The second panel, Fig.~\ref{fig:max-order}(b),
shows the dependence of the error for the VCP method depending on
the contour radius $r$, justifying the choice of $r=0.1$.

\begin{figure}[t]
\begin{centering}
\includegraphics[width=0.9\columnwidth]{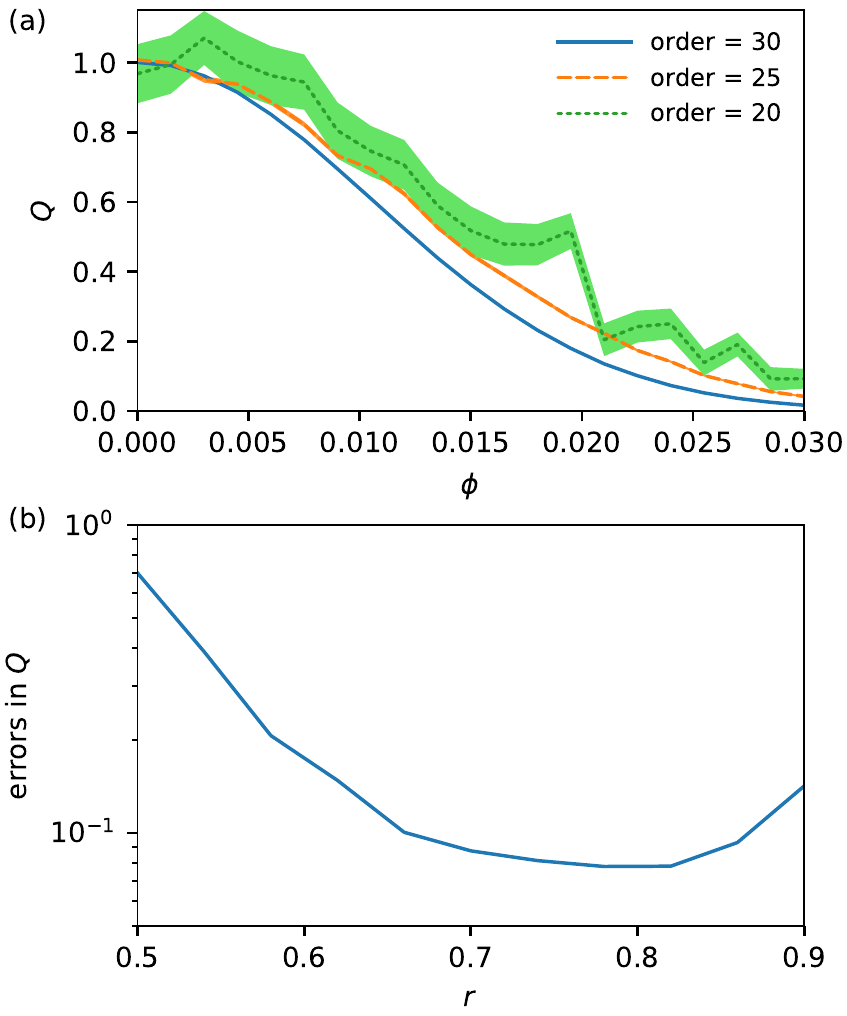}
\par\end{centering}
\caption{\label{fig:low-order}(a) Correlations of different order for $M=30$
using the VCP method with $200$ ensembles of $10^{6}$ samples. The
VCP radius was was $r=0.8$ to minimize sampling errors (see panel
(b)). (b) Dependence of the error for the $25$-th order correlation
as a function of $r$ for $M=30$ and $\phi=0.01$, using VCP with
$200$ ensembles of $10^{4}$ samples. }
\end{figure}

Next, we include the effects of decoherence by using an independent phase
noise term $\xi$ at each site with a normal distribution, having
zero mean and variance $\sigma^{2}$. This is important, since many
interferometry proposals are fragile against decoherence. Each mode
has a noise term given by $\phi_{j}=j\phi+\xi_{j}$, $j=1,\dots,M$.
This effect is shown in Fig~\ref{fig:noisy} for different values
of the variance. 

We also evaluate different order correlations, since experimental
detector inefficiencies may lead to greater count rates for lower
order correlations. This is shown in figure~\ref{fig:low-order}(a),
for $M=30$, where we show three different order correlation functions,
for $Q=\left\langle \hat{n}_{1}\ldots\hat{n}_{N}\right\rangle $,
with $N=30$, $N=25$ and $N=20$. As the order is reduced the fringe
width does not change dramatically. The interferometer is robust against lower
order measurements.

Figure~\ref{fig:low-order}(b) gives the sampling error as a function
of $r$. This demonstrates how the choice of integration contour can
change the sampling error for these lower order correlation measurements.
The results show that in this case, one should use a contour near
$r=1$. However, the exact choice of radius is important. For this
order of correlation ($N=25$), we find that for the lower-order correlations
there is a finite optimal value of $r$, in this case near $r=0.8$,
where the error is the lowest. 

In summary, we have developed two probabilistic methods based on complex
$P$-distributions, that can be used to find the expectation values
of the correlations of the outputs of a Boson Sampling interferometer.
The general one, VCP, handles correlations of any order. The more
specialized QCP is limited to the correlations of maximum order (equal
to the number of input photons), but has a much lower sampling error
than VCP\@. These results are limited by sampling errors. The advantage
of sampled contour integration is that the complex P-distribution
has a compact support, reducing sampling error. This approach is related to
  contour integrals for matrix permanents~\cite{fyodorov2006permanental}. 

We used these methods to simulate QuFTI experiments with up to $100$
modes, under noisy conditions, by adding noise to the measured phase
gradient and considering lower order correlations. Our results demonstrate an exceptional robustness of multi-photon
quantum Fourier interferometry against decoherence. These methods
may have a wider applicability in analyzing decoherence in quantum
photonic networks, and to related problems in quantum information.

\bibliography{qufti_letter}

\end{document}